\pdfoutput = 1
\documentclass[%
 reprint,
 prd,
 amsmath,amssymb,
 aps,
 superscriptaddress,
 floatfix
]{revtex4-1}

\usepackage{dcolumn}
\usepackage{bm}
\usepackage{ulem}
\usepackage{multirow}
\usepackage{placeins}
\usepackage{algorithmic}
\usepackage{color}
\usepackage{comment}

\usepackage{array}
\usepackage{hyperref}
\newcolumntype{P}[1]{>{\centering\arraybackslash}p{#1}}
\usepackage{xcolor}

\begin{document}
\title{Inaccessibility of the Schwarzschild Singularity to Quantum Probes}
\author{Vishnulal Cheriyodathillathu}\thanks{lal23@iisertvm.ac.in}
\affiliation{School of Physics, Indian Institute of Science Education and Research Thiruvananthapuram, Maruthamala PO, Vithura, Thiruvananthapuram 695551, Kerala, India}
\author{Tanmay Patil}\thanks{patiltn2601@gmail.com}
\affiliation{Department of Physics, Imperial College London, London,UK}
\author{Saurya Das}\thanks{saurya.das@uleth.ca}
\affiliation{Quantum Horizons Alberta \& Theoretical Physics Group, Department of Physics and Astronomy, University of Lethbridge 4401 University Drive, Lethbridge, Alberta T1K 3M4, Canada}
\author{Soumen Basak}\thanks{sbasak@iisertvm.ac.in}
\affiliation{School of Physics, Indian Institute of Science Education and Research Thiruvananthapuram, Maruthamala PO, Vithura, Thiruvananthapuram 695551, Kerala, India}

\begin{abstract}
We revisit the Schwarzschild singularity in a semiclassical setting where the background geometry is classical and quantum effects enter through Bohmian (quantal) trajectories associated with a Klein-Gordon wave function. Using the Madelung-Bohm decomposition of the Klein-Gordon wavefunction, we show that the quantum-modified motion is equivalent to geodesic motion in an effective metric conformally related to Schwarzschild, with a conformal factor fixed by the wavefunction amplitude. Solving the wavefunction equation near $r\to 0$ determines this factor and yields finite curvature invariants, in suitable coordinates the interior extends smoothly and the effective spacetime is geodesically complete. Our result suggests that quantum probes on a prescribed classical background can experience an operationally regular effective geometry, even though the underlying Schwarzschild spacetime itself remains classically singular.
\end{abstract}

\maketitle

\setlength{\columnsep}{0.5cm}

\section{Introduction}

General relativity predicts that, under a broad set of conditions, classical spacetimes possess singularities. In many cases, curvature invariants, such as the Kretschmann scalar, diverge. However, as established by the Hawking-Penrose singularity theorems, the definitive criterion is the termination of timelike or null geodesics at a finite value of proper time or affine parameter, i.e., geodesic incompleteness \cite{HawkingPenrose1970,Penrose1965}. It is widely expected that a complete quantum theory of gravity will resolve these pathologies and yield a non-singular description of black holes and cosmological spacetimes. Indeed, approaches ranging from string theory \cite{Polchinski1998,Zwiebach2009}, loop quantum gravity \cite{Rovelli2004,AshtekarLewandowski2004}, asymptotic safety \cite{Reuter1998,Percacci2009} and studies of gravitational collapse using a functional Schr\"odinger formalism \cite{EricGreenwood_2008,PhysRevD.89.044003} suggest modifications of classical geometry near Planck-scale curvatures that could enable singularity resolution.

Following the procedure in \cite{PhysRevD.89.084068}, a complementary perspective comes from the Raychaudhuri equation \cite{PhysRev.98.1123} and its quantum extension, the quantum Raychaudhuri equation. In this framework, classical geodesics are replaced by Bohmian (quantal) trajectories \cite{PhysRev.85.166,PhysRev.85.180} associated with an underlying quantum wavefunction. In the Bohmian formulation, the phase of the wavefunction defines a local momentum field, whose integral curves provide a trajectory representation of the corresponding quantum phase flow. A ``repulsive'' quantum term then appears, counteracting the usual ``attractive'' focusing and, in principle, preventing the formation of conjugate points along the worldlines \cite{PhysRevD.89.084068,ALI2016173}. The resulting trajectories obey deterministic equations of motion.
These quantum corrections may be equivalently expressed as an effective force that drives deviations from classical geodesic motion and captures the influence of the underlying quantum wavefunction. In this work, a quantum probe denotes a Klein–Gordon test field configuration on the prescribed Schwarzschild background together with the local phase flow defined by its phase. We show that the resulting quantal trajectories admit a natural geometric reinterpretation as geodesics of a conformally related spacetime. In this formulation, the effects of the quantum force are entirely absorbed into an appropriate conformal factor, providing a purely geometric description of the modified dynamics experienced by the quantum probe. 
This geometric reformulation also suggests a natural and physically motivated criterion for singularity resolution in the effective spacetime. One must establish not only that the curvature invariants remain finite but also that the geodesics governing the probe dynamics are complete in the effective spacetime. We therefore investigate the singularity structure of the resulting conformally related spacetime from both perspectives. Our analysis indicates that the effective geometry experienced by the associated phase flow lines can be free of curvature singularities, suggesting a possible mechanism for the regularization of classical black hole interiors within this framework. This operational viewpoint is analogous in spirit to fixed background quantum probe approaches to spacetime singularities. For example, Horowitz and Marolf proposed that a classically singular spacetime may be quantum mechanically nonsingular if the evolution of a quantum test state is uniquely defined on the prescribed singular background \cite{PhysRevD.52.5670}. In this work, we ask whether the phase flow lines associated with selected Klein-Gordon probe modes on the prescribed Schwarzschild background reach the classical $r=0$ singular surface in finite effective proper time. Thus, the singularity regularization discussed here should be understood as an operational statement.

The remainder of the paper is organized as follows. In Section~II, we derive the quantum-modified geodesic equation from the Klein-Gordon wavefunction using the Madelung-Bohm approach. In Section~III, we use the idea of conformal transformation to reinterpret the quantum modified motion as geodesics in a conformally related geometry. We extend this approach to a more general class of Klein-Gordon equations. In Section~IV, we solve the Klein-Gordon equation on Schwarzschild background to construct the conformally related geometry and thereby calculate its associated curvature invariants. In Section~V, we introduce an appropriate coordinate system in which the regularity of the conformally related geometry becomes manifest. Within this framework, we further examine the global structure of the spacetime and demonstrate geodesic completeness of the effective geometry, showing consistency with the general picture of quantum modified motion as geodesic flow in a conformally related spacetime. In Section~VI, we summarize our main results and discuss possible extensions and implications of the present approach.

\section{Quantum Corrections To The Geodesic Equation}
 
The central idea of this work is to geometrically reformulate the quantum modified dynamics of the probe, so that the corresponding Bohmian phase flow lines become geodesics of a suitably defined effective metric. We apply the Madelung-Bohm decomposition to the Klein-Gordon equation in curved spacetime. Throughout, we assume a fixed classical background with Lorentzian signature $(+,\,-,\,-,\,-)$ and work in units with $c=1$ unless otherwise stated. Our analysis begins with the Klein-Gordon equation for a massive scalar field $\Phi$,
\begin{equation}
     \Box \Phi + \dfrac{m^2}{\hbar^{2}}\Phi=0.
    \label{eqm}
\end{equation}
Here, $\Box\!\equiv \!g^{\mu\nu}\nabla_\mu\nabla_\nu$ denotes the covariant d’Alembertian associated with the background metric $g_{\mu\nu}$. To define a quantum velocity field, we write
\begin{equation}
   \Phi(\vec{x},t) = \mathcal{R}e^{i\mathcal{S}}~, 
    \label{ansatz}
\end{equation}
where $\mathcal{R}(\vec{x},t)$ and $\mathcal{S}(\vec{x},t)$ are real functions. Substituting Eq.~\eqref{ansatz} into Eq.~\eqref{eqm} and separating the real and imaginary parts gives
\begin{equation}
     \nabla_{\mu}\left(\mathcal{R}^{2} \partial^{\mu} \mathcal{S}\right)\,=\,0,
     \label{im}
\end{equation}
\begin{equation}
    \dfrac{\Box \mathcal{R}}{\mathcal{R}} - \left(\partial^{\mu}\mathcal{S}\right)\left(\partial_{\mu}\mathcal{S}\right)=-\dfrac{m^{2}}{\hbar^{2}}~.
    \label{HJ}
\end{equation}
The real part may be recast as
\begin{equation}
    p_{\mu} p^{\mu} = \mathcal{M}^{2},
    \label{effective mass}
\end{equation}
where
\begin{equation}
    \mathcal{M}^{2} =  m^{2} + \hbar^{2}  \dfrac{\Box \mathcal{R}}{\mathcal{R}}. \label{Msquared}
\end{equation}
and the four momentum $p_{\mu}$ and the four velocity $U_{\mu}$ are defined as follows \cite{10.1143/ptp/90.4.753,GeorgeHorton_2000,10.1098/rspa.2013.0699,HrvojeNikolic2007},
\begin{equation}
    p_{\mu} = \hbar \nabla_{\mu}\mathcal{S},
\end{equation}
\begin{equation}
    U_{\mu} = \dfrac{p_\mu}{\mathcal{M}},  \hspace{30pt} U_{\mu} U^{\mu} = 1~.  \label{norm}
\end{equation}
The vector field $p^\mu$ therefore defines a local phase flow, whose integral curves constitute the Bohmian type trajectories considered in this work. Eq.~\eqref{effective mass} is a quantum-modified dispersion relation. The normalized four-velocity $U^{\mu}$ then satisfies the quantum-modified geodesic equation
\begin{equation}
    U^{\mu}\nabla_{\mu}U^{\alpha} = \dfrac{\hbar^{2}}{2 \mathcal{M}^{2}} h^{\mu \alpha} \nabla_{\mu} \left(\dfrac{\Box \mathcal{R}}{\mathcal{R}}\right), \label{modifiedgeodesic}
\end{equation}
where
\begin{equation}
    h^{\mu \alpha} = g^{\mu \alpha} - U^{\mu} U^{\alpha}.
    \label{proj}
\end{equation}
The non-vanishing right-hand side originates from the quantum nature of the probe and depends explicitly on its wavefunction. It acts as an effective force, causing deviations from the classical geodesics of the background metric $g_{\mu \nu}$, although the spacetime geometry itself is held fixed and classical. The projection tensor $h_{\mu\alpha}$ ensures that this quantum force is orthogonal to the four-velocity $U^{\mu}$, thus preserving the normalization condition in Eq.~\eqref{norm} along the modified trajectory \cite{Shojai:1998qou}. The classical geodesic limit is recovered when the quantum potential contribution becomes negligible.
%
\section{Quantum corrected geometry}
The modified quantum geodesic equation (\ref{modifiedgeodesic}) helps us to interpret the motion of a quantum probe in spacetime in a geometric way \cite{Broglie1960NonlinearWM}. First, the above quantum modified geodesic equation can be derived from an action principle by replacing the mass term in the action for a relativistic particle with the effective mass $\mathcal{M}$, as defined in Eq.(\ref{Msquared}).
\begin{equation}
    \mathcal{A} = -\int \mathcal{M}(x) ds = -m\int\left(1+\dfrac{\hbar^{2}}{m^{2}}\dfrac{\Box{\mathcal{R}}} {\mathcal{R}}\right)^{1/2} \sqrt{g_{\mu \nu} dx^{\mu} dx^{\nu}}.
    \label{action1}
\end{equation}
Second, the modified dispersion relation can be written as
\begin{equation}
   \dfrac{1}{\Omega^{2}} g^{\mu \nu} p_{\mu} p_{\nu} = m^{2},
   \label{moddisp}
\end{equation}
where
\begin{equation}
    \Omega^{2} = \left(1+\dfrac{\hbar^{2}}{m^{2}}\dfrac{\Box{\mathcal{R}}} {\mathcal{R}}\right) = \dfrac{\mathcal{M}^{2}}{m^{2}}.
    \label{omega}
\end{equation}
Equations~\eqref{action1} and \eqref{moddisp} suggest that we can interpret quantal trajectories as geodesics in a conformally related metric \cite{doi:10.1142/S0217751X98000305,article,doi:10.1142/S0217732319502705,Chowdhury:2021fjw},
\begin{equation}
    \Tilde{g}_{\mu \nu} = \Omega^{2} g_{\mu \nu}~. \label{gtilde}
\end{equation}
This $\Tilde{g}_{\mu \nu}$ is the effective state-dependent metric, as seen by the quantum probe in the background spacetime. The corresponding equation of motion can be written as
\begin{equation}
    \Tilde{U}^{\mu}\Tilde{\nabla}_{\mu}\Tilde{U}^{\alpha} = 0~,
    \label{gbargeo2}
\end{equation}
where $\Tilde{U}^{\mu}$ is the normalized four-velocity in the $\Tilde{g}$ metric. The normalized four-velocities in the spacetimes $g$ and $\Tilde{g}$ are related by,
\begin{equation}
    \Tilde{U}^{\mu} = \dfrac{1}{\Omega} U^{\mu}.
\end{equation}
Now we have two spacetime geometries, $g$ and $\Tilde{g}$, and the non-geodesic motion in the original background metric can be re-expressed as geodesic motion in the conformally related metric for the class of probes being considered. The effective metric interpretation is meaningful in the admissible region of the probe configuration space where $\mathcal{M}^{2}>0$. Therefore, the positivity of $\mathcal{M}^2$ is not assumed for arbitrary Klein-Gordon modes. Rather, it is an admissibility condition for the Bohmian effective-metric construction used in this work. For $\mathcal M^2<0$, the phase momentum field becomes spacelike, a kinematic situation related to tachyonic motion \cite{Bilaniuk1962,Bilaniuk1969}, although here $\mathcal M^2$ is only a local, state dependent effective quantity and does not by itself imply a tachyonic excitation.

To examine the applicability of the effective geometric construction beyond the minimally coupled Klein-Gordon equation, we consider a broader class of scalar probe equations incorporating an additional local covariant scalar coupling. From an effective field theory perspective, curvature dependent scalar terms provide a natural class of covariant corrections to the probe dynamics. We therefore consider,
\begin{equation}
    \big(\Box + \dfrac{m^{2}}{\hbar^{2}}+ \eta \,\,I(x) \big) \Phi = 0~,
    \label{coKG}
\end{equation}
where $I(x)$ is a scalar constructed from the background geometry or other local spacetime fields, and $\eta$ is a positive coupling constant chosen so that $\eta I(x)$ has dimensions of inverse length squared. This equation should be regarded as an effective potential-type modification of the Klein-Gordon equation rather than as the most general EFT completion, since derivative and higher-order self-interaction terms are not included. Using the ansatz in Eq.~\eqref{ansatz} and following the same procedure as in the previous section, we obtain the modified dispersion relation
\begin{equation}
    g^{\mu \nu} p_{\mu} p_{\nu} = m^{2} \big(1 + \dfrac{\hbar^{2}}{m^{2}} \dfrac{\Box{\mathcal{R}}} {\mathcal{R}} +  \dfrac{\hbar^{2}}{m^{2}} \eta I \big) = \mathcal{M}^{2}
    \label{cmod}
\end{equation}
This can be recast into the $\Tilde{g}$ background as follows,
\begin{equation}
    \Tilde{g}^{\mu \nu} p_{\mu} p_{\nu} = m^{2}~,
    \label{cmoddisp}
\end{equation}
where
\begin{equation}
\begin{aligned}
    \Tilde{g}_{\mu \nu} &= \Omega^{2} g_{\mu \nu} \, ,\\
    \Omega^{2} &= \bigg(1 + \dfrac{\hbar^{2}}{m^{2}} \dfrac{\Box{\mathcal{R}}}{\mathcal{R}} +  \dfrac{\hbar^{2}}{m^{2}} \,\eta I \bigg) ~.
\end{aligned}
\end{equation}
Now, the four velocities associated with these quantal trajectories follow an equation similar to \eqref{gbargeo2} with modified factors. 
The modified dispersion relation \eqref{cmod} shows that the quantum potential and curvature coupling combine into an effective mass function $\mathcal{M}$, so that the corresponding quantal trajectories, though non-geodesic in the original spacetime $g_{\mu \nu}$, are precisely geodesics of a conformally related metric $\Tilde{g}_{\mu \nu}$.

In a general curved spacetime, a preferred global positive frequency decomposition, and hence a preferred particle interpretation, need not exist. Moreover, for a general Klein-Gordon solution, the time component of the conserved Klein-Gordon current need not be positive definite and therefore cannot, in general, be interpreted as a probability density. For the present construction, however, a globally defined positive frequency decomposition is not required. Wherever $\mathcal R\neq0$, the current is proportional to $\mathcal R^2p^\mu$ and follows the same unparametrized curves as $p^\mu$. In regions where $\mathcal M^2>0$, these curves admit a timelike geodesic interpretation in the effective conformal metric. The usual interpretational subtleties of relativistic Klein-Gordon/Bohmian mechanics therefore remain, but they do not affect the local effective metric construction used here, which depends only on the phase flow structure of the selected Klein-Gordon field configuration \cite{PhysRevLett.93.090402,Nikolic:2004tu,Alkhateeb:2022vcf}. More recent works have further developed geometric and operational formulations of relativistic Bohmian motion \cite{Foo_2022, PhysRevA.111.052201, hatifi2025quantumgravitymetricquantization, LANTIGUA2025170226}.

The conformal metric thus provides the relevant geometry for probing spacetime at the quantum level, allowing the nature of black hole singularities to be assessed in terms of the metric governing quantum motion. This motivates the analysis of curvature invariants of $\Tilde{g}_{\mu \nu}$, which we carry out in the next section to examine the regularity of the effective geometry near the core.
%
\section{Curvature scalars in conformally related metric}

From this section onward, we choose Schwarzschild spacetime as our background metric ($g_{\mu \nu}$). Quantal trajectories can be interpreted as geodesics of a $\Tilde{g}_{\mu \nu}$ metric conformally related to the Schwarzschild metric, with the conformal factor determined by the selected Klein-Gordon field configuration and the local potential. Our objective is to study the regularity of the effective metric $\Tilde{g}_{\mu \nu}$. For that purpose, we analyze the  $r\rightarrow 0$ limit of the Kretschmann scalar in the $\Tilde{g}_{\mu \nu}$ metric,
\begin{equation}
    \Tilde{\mathcal{K}} = \Tilde{R}^{\mu \nu \lambda \rho} \Tilde{R}_{\mu \nu \lambda \rho} = \Tilde{C}^{\mu \nu \lambda \rho} \Tilde{C}_{\mu \nu \lambda \rho} + 2 \Tilde{R}^{\mu \nu} \tilde{R}_{\mu \nu} - \dfrac{1}{3} \Tilde{R}^{2}~,
    \label{kret}
\end{equation}
where $\Tilde{C}_{\mu \nu \lambda \rho}$ is the Weyl tensor, $\Tilde{R}_{\mu \nu}$ the Ricci tensor, and $\Tilde{R}$ the Ricci scalar. We define
\begin{equation}
    \psi \equiv \dfrac{1}{2} \ln{\Omega^{2}} \,\,\,\,\,  \Rightarrow \,\,\,\,\, \Tilde{g}_{\mu \nu} = e^{2 \psi} g_{\mu \nu}~.
\end{equation}
Under a conformal transformation of this form, the quantities in Eq.~\eqref{kret} transform as \cite{Wald:1984rg},
\begin{equation}
    \Tilde{C}_{\mu \nu \lambda \rho} = e^{2\psi} C_{\mu \nu \lambda \rho},
\end{equation}
\begin{equation}
 \Tilde{R}_{\mu \nu}=R_{\mu \nu}-2\nabla_{\mu}\nabla_{\nu}\psi-g_{\mu \nu} \Box{\psi}+2\nabla_{\mu}\psi \nabla_{\nu}\psi-2g_{\mu \nu}(\nabla \psi)^{2},
\end{equation}
\begin{equation}
    \Tilde{R} = e^{-2\psi} \left[R-6\Box{\psi}-6(\nabla \psi)^{2}\right].
\end{equation}
For Schwarzschild spacetime $R_{\mu \nu}=R=0$, and this simplifies the above expressions. We now write an ansatz for the asymptotic form of $\Omega$ as,
\begin{equation}
    \Omega^{2}(r) \sim r^{-n} \,\,\,\,\,\,\,\,\,\, \text{when}\,\,\,\,\,\,  r \rightarrow 0.
    \label{confo}
\end{equation}
Here, $n>0$ characterizes the leading power-law growth of the conformal factor as $r\to0$. This together with the above transformation equations yields the behavior of the Kretschmann scalar in the $\Tilde{g}_{\mu \nu}$ metric near the classical singularity,
\begin{equation}
    \Tilde{\mathcal{K}} \sim r^{2n-6} \,\,\,\,\,\, \text{when} \,\,\,\, r\rightarrow0.
    \label{ktil}
\end{equation}
Similarly, we can also calculate how the Ricci scalar and the square of the 
Ricci tensor behave near the singularity in the $\tilde{g}_{\mu \nu}$ metric,
\begin{equation}
    \Tilde{R} \sim r^{n-3}   \hspace{20pt} \Tilde{R}_{\mu \nu}\Tilde{R}^{\mu \nu}\sim r^{2n-6}
\end{equation}
Thus, for $n\ge 3$ we have finite curvature invariants as $r\to 0$ in the $\Tilde{g}_{\mu \nu}$ metric. In fact, the number 3 signifies the number of spatial dimensions in the spacetime geometry we have chosen. This is easy to see if we use the Schwarzschild Tangherlini black hole which is the generalization of the Schwarzschild black hole to $(d+1)$- spacetime dimensions \cite{Tangherlini:1963bw,Singh:2017vfr}. Then the corresponding calculations will show that the curvature scalars are finite whenever $n \ge d$, the number of spatial dimensions.

Now we show that such a conformal factor comes out naturally. We consider a massive Klein-Gordon field propagating in a Schwarzschild black hole background, starting from the more general version,
\begin{equation}
   \big( \Box{} + \dfrac{m^{2}}{\hbar^{2}} + \xi R + \eta \hbar G \mathcal{K} \big)\Phi =0~,
   \label{coKG1}
\end{equation}
where $\xi$ and $\eta$ are dimensionless coupling constants. The metric of the black hole geometry is as follows,
\begin{equation}
    ds^{2} = f(r)dt^{2} - \dfrac{1}{f(r)} dr^{2} - r^{2} dL_{\perp}^{2}, \,\,\,\, f(r) = 1 - \dfrac{r_{s}}{r} ~.
\end{equation}
Here, $dL_{\perp}^{2}$ denotes the metric on the $t=\text{constant}$ and $r=\text{constant}$ hypersurface and $r_{s}$ is the Schwarzschild radius. For the case of Schwarzschild black hole the Ricci scalar vanishes and Eq.~\eqref{coKG1} reduces to Eq.~\eqref{coKG} with $I$ replaced by $\hbar G \mathcal{K}$. We begin with the following ansatz for $\Phi$,
\begin{equation}
   \Phi(r,t) = \mathcal{R}(r)e^{i\mathcal{S}(r,t)}~, \,\,\,\,\,\,\, \mathcal{S}(r,t) = -k t + W(r) ~.
    \label{ansatz1}
\end{equation}
Following a similar procedure as in Section II, we can split the Klein-Gordon equation into imaginary and real parts. The imaginary part is the continuity equation that ensures current conservation.
\begin{equation}
    \nabla_{\mu}(\mathcal{R}^{2}g^{\mu \nu} \partial_{\nu}\mathcal{S}) = 0 \,\,\,\Rightarrow \,\,\,  r^{2}\mathcal{R}^{2}f(r)W^{'}(r) = D
\end{equation}
Here $D$ is a positive constant corresponding to inward directed flux. This means that the flux is towards decreasing $r$. The real part is the differential equation for the amplitude $\mathcal{R}(r)$ and reduces to the following form in the limit $r\to0$.
\begin{equation}
    \mathcal{R}^{''} + \dfrac{1}{r} \mathcal{R}^{'} + \left[\dfrac{-D^{2}}{r^{4} f^{2} \mathcal{R}^{4}} + \dfrac{\beta}{r^{5}}\right] \mathcal{R} \simeq 0,
    \label{33}
\end{equation}
where
\begin{equation}
    \beta =  12  \eta  \hbar G {r_{s}}~.
\end{equation}
To obtain the leading asymptotic solution of Eq.~\eqref{33}, we take the power law ansatz, $\mathcal R(r)=A r^q $. Using $f(r)\simeq-r_s/r$ as $r\to0$, the dominant terms in Eq.~\eqref{33} scale consistently for $q=3/4$, while their coefficients give $A^4=\frac{D^2}{\beta r_s^2}$. The final solution for the field $\Phi(r,t)$ is obtained as follows,
\begin{equation}
    \Phi(r,t) = \left(\dfrac{D^{2}}{\beta r_{s}^{2}}\right)^{\dfrac{1}{4}} r^{\dfrac{3}{4}}\exp{\left[i(-k t + \dfrac{2}{3} \sqrt{\beta} r^{-\dfrac{3}{2}})\right]}, \,\,\, \eta>0
\end{equation}
Using this to construct the conformal factor gives
\begin{equation}
    \Omega^{2} = \left(1+\dfrac{\hbar^{2}}{m^{2}}\dfrac{\Box{\mathcal{R}}} {\mathcal{R}} + \dfrac{\hbar^{2}}{m^{2}} \eta \hbar G \mathcal{K}\right) \sim r^{-6}
    \label{Omegafinal}  \,\,\,\,\,\,\,\text{when} \,\,\,r\rightarrow0.
\end{equation}
For the explicit solution considered here, the positive curvature coupling contribution dominates asymptotically, yielding $\Omega^2\sim r^{-6}>0$ for $\eta>0$, and hence $\mathcal{M}^2=m^2\Omega^2>0$ as $r\to0$. This timelike character is a property of the selected solution and is not imposed on arbitrary curves in the Schwarzschild interior.
Comparing Eq.~\eqref{Omegafinal} with Equations~\eqref{confo} and \eqref{ktil} immediately shows that $\Tilde{\mathcal{K}}$ remains finite as $r\to0$. Using the probe wave function, we explicitly construct the conformal factor that defines the effective metric $\Tilde{g}_{\mu \nu}$. The resulting geometry is regular in the Schwarzschild interior, and a direct evaluation of curvature invariants shows that the Kretschmann scalar associated with $\Tilde{g}_{\mu \nu}$ remains finite in the $r\to0$ limit. Since $\mathcal{M}^{2}$ remains positive as $r\to0$, the phase momentum $p_{\mu}=\hbar\nabla_{\mu}\mathcal{S}$ continues to satisfy a timelike mass shell condition in the effective geometry, allowing the associated Bohmian phase flow curves to retain a timelike trajectory interpretation. Thus, the effective geometry experienced by the selected probe configuration is curvature-regular in the near core limit.

We can generalize this to a broader class of Klein-Gordon extensions of the form Eq.~\eqref{coKG} with $I \sim r^{-n}$ with $n\ge3$ as $r\to0$. In these cases, the term $\dfrac{\Box{\mathcal{R}}} {\mathcal{R}}$ in the expression of the conformal factor will be subleading compared to the coupling term when $n>3$ and vanishes when n=3.
This suggests that the conformal factor is driven by the coupling factor in the $r\to0$ limit. Thus, whenever we have a curvature coupled Klein-Gordon equation with $I \propto r^{-n}$ and $n\ge3$, the Kretschmann scalar is finite at the singularity. 

The energy density of the selected Klein-Gordon probe generally grows as the classical $r=0$ singular surface is approached, which can limit the validity of fixed background approximation used in our work. However, this approximation can be controlled by an appropriate constant rescaling of the probe mode, $\Phi(x) \longrightarrow \Phi_{\epsilon}(x) = \epsilon \Phi(x)$. Under this transformation, $\Omega^{2}$ remains invariant, while the energy momentum tensor $T_{\mu \nu}$ scales quadratically with the overall field amplitude parameter $\epsilon$. Thus, on any fixed cutoff domain away from $r=0$, the probe's energy density can be made arbitrarily small while preserving the effective geometry experienced by the probe. For completeness, the detailed derivation of this test field scaling and the corresponding backreaction parameter is provided in the Appendix.

\section{Geodesic Completeness}
%
The finiteness of the Kretschmann scalar $\Tilde{\mathcal{K}}$ as $r\to0$ does not by itself guarantee that the effective metric $\Tilde{g}_{\mu \nu}$ is free from curvature singularities, since it is not a sufficient condition for establishing the complete regularity of the spacetime. A stronger notion of regularity also requires geodesic completeness so that causal geodesics do not terminate after a finite affine parameter. In the radial coordinate $r$, the metric components remain singular as the value of $r$ tends to $0$, which hides the geometric interpretation of this $r=0$ surface. Therefore, we must further determine whether the surface $r=0$ represents a genuine curvature singularity, a removable coordinate singularity, or an asymptotic boundary located at an infinite geodesic distance. To analyze the geometric nature of this surface, we begin with the conformally related metric,
\begin{equation}
    d\Tilde{s}^{2} = \Omega^{2} \left(f(r) dt^{2} - \dfrac{1}{f(r)} dr^{2} - r^{2} dL^{2}_{\perp}\right), \label{ds2}
\end{equation}
where
\begin{equation}
    f = 1-\dfrac{r_{s}}{r} \sim -\dfrac{r_{s}}{r} \,\,\,\,\,\,\,\,\, \text{when}\,\,\,\, r\rightarrow0,
\end{equation}
and $dL^{2}_{\perp}$ is the angular part of the metric, we use Eq.~\eqref{confo} in the form,
\begin{equation}
    \Omega^{2} = C r^{-n}.
    \label{e3}
\end{equation}
Inside the event horizon, space and time swap roles, so $r$ is timelike. Keeping this in mind, to disentangle the above possibilities, we introduce a proper time coordinate $\Tilde{\tau}$ defined as follows,
\begin{equation}
\Tilde{\tau} =
\begin{cases}
\frac{2}{n-3} \sqrt{\frac{C}{r_{s}}}\,\left(r^{\frac{3-n}{2}} - r_{0}^{\frac{3-n}{2}} \right) \,\,\,\, & \text{for $n\neq3$} \\[14pt]
\sqrt{\frac{C}{r_{s}}} \ln{\frac{r}{r_{0}}} \,\,\,\,  & \text{for  $n=3$}
\label{tau}
\end{cases}
\end{equation}
In either case, the metric becomes
\begin{equation}
    d\Tilde{s}^{2} = d\Tilde{\tau}^{2} - a(\Tilde{\tau})^{2} dt^{2} - b(\Tilde{\tau})^{2} dL_{\perp}^{2},
\end{equation}
where
\begin{equation}
a(\Tilde{\tau})^{2} =
\begin{cases}
a_{0} \Tilde{\tau}^{\dfrac{2(1+n)}{n-3}}  \,\,\,\, & \text{for $n \ne 3$} \\[14pt]
\bar{a}_{0} e^{-4\sqrt{\dfrac{r_{s}}{C}}\Tilde{\tau}} \,\,\,\,& \text{for $n=3$}
\end{cases}
\end{equation}
\begin{equation}
b(\Tilde{\tau})^{2} =
\begin{cases}
b_{0} \Tilde{\tau}^{\dfrac{2(n-2)}{n-3}} \,\,\,\, & \text{for $n \ne 3$} \\[14pt]
\bar{b}_{0} e^{-\sqrt{\dfrac{r_{s}}{C}}\Tilde{\tau}} \,\,\,\, & \text{for $n=3$}
\end{cases}
\end{equation}
Although the metric coefficients diverge in the original $r$-coordinate system, the $\Tilde{\tau}$-coordinate representation shows that this divergence occurs only at  $|\Tilde{\tau}|\to \infty$ whenever $n\ge3$. This means that the singular surface $r=0$ is an asymptotic boundary rather than a finite coordinate location. The divergence of $\Tilde{\tau}$ as $r \to 0$ directly reflects the fact that causal geodesics require infinite affine/proper time parameter to reach this surface. Thus, for $n\ge3$, the conformally related spacetime is geodesically complete toward the center.
\subsection{Geodesics in the Conformally Related Schwarzschild Geometry}
We now analyze radial timelike geodesics in the conformally modified Schwarzschild metric as a consistency check of the geodesic completeness derived from the interior coordinate construction. We start from the following form of the metric,
\begin{equation}
d\tilde{s}^2 = h_1(r)dt^2 - h_2(r)dr^2 - r^2\Omega^2(r)dL^{2}_{\perp}~,
\end{equation}
where
\begin{equation}
h_1(r)=\Omega^2(r)f(r), \qquad
h_2(r)=\Omega^2(r)f^{-1}(r),
\end{equation}
For time like geodesics, the four velocity satisfies the normalization condition,
\begin{equation}
\tilde g_{\mu\nu}\tilde U^\mu \tilde U^\nu = 1 .
\end{equation}
For purely radial motion, $\tilde U^2=\tilde U^3=0$. The zeroth component of the geodesic equation yields
\begin{equation}
\dfrac{d\tilde U^0}{d\Tilde{\tau}} + 2\Tilde{\Gamma}^{0}_{01}\tilde U^0 \tilde U^1 =0 ,
\end{equation}
which implies the conserved quantity,
\begin{equation}
h_1(r)\tilde U^0 = \tilde{E} .
\end{equation}
Using this result together with the normalization condition we obtain the following result,
\begin{equation}
(\tilde U^1)^2
=\dfrac{1}{h_1(r)h_2(r)}
\left[\tilde{E}^2-h_1(r)\right].
\end{equation}
Substituting the behavior of the near origin $\Omega^2(r)\propto r^{-n}$ and $f(r)\sim -\dfrac{r_s}{r}$, we find the following.
\begin{equation}
(\tilde U^1)^2 \propto r^{n-1}, 
\end{equation}
Then the proper time taken by a radial timelike geodesic to reach the singular $r=0$ surface can be calculated as
\begin{equation}
\Tilde{\tau} \sim \int_{0}^{\epsilon} r^{\frac{1-n}{2}} dr
\end{equation}
This integral diverges for $n\ge3$, consistent with Eq.(\ref{tau}). One can also use the Killing vector $(\partial_t)^{\mu}$ associated with the Schwarzschild geometry to obtain the conserved quantity and arrive at the same condition for $n$. 

A similar analysis can be performed for null radial geodesics. Under conformal transformations, null curves remain as null curves. This means that the path of a null geodesic is unchanged by the conformal transformation, but the affine parameter changes. If $\lambda$ is the affine parameter of $g_{\mu \nu}$, then the affine parameter $\Tilde{\lambda}$ for the conformal metric satisfies the following relation,
\begin{equation}
    d\Tilde{\lambda} = \Omega^{2} d\lambda~.
    \label{e1}
\end{equation}
We start from the null condition for radial null geodesics in $g_{\mu \nu}$ metric,
\begin{equation}
    0=f(r) \Dot{t}^{2} - \frac{1}{f(r)} \Dot{r}^{2}~,
    \label{e2}
\end{equation}
where the dots represent derivatives with respect to the affine parameter $\lambda$. As we mentioned earlier, even though the $t$ coordinate changes its role inside the event horizon, the geometry still admits a Killing vector $(\partial_{t})^{\mu}$. This helps us to write a conserved quantity, $E=f(r) \Dot{t}$. This fact together with Equations (\ref{e1}), (\ref{e2}) and (\ref{e3}) gives the following relation for $\Tilde{\lambda}$ in the $r$ tends to zero limit,
\begin{equation}
    \Tilde{\lambda} = \int_{0}^{\epsilon} \frac{C}{E} r^{-n} dr ~.
\end{equation}
The above integral diverges for $n\ge1$. We can even extend this calculation to the case of non-radial null geodesics. This will add another conserved quantity, $L=r^{2}\Dot{\phi}$, which is associated with the Killing vector $(\partial_{\phi})^{\mu}$. The corresponding calculation gives us the geodesic completeness for $n\ge\frac{5}{2}$. So for our case, $n\ge3$ the null geodesics reach $r=0$ only after an infinite conformal affine parameter. We have already seen geodesic completeness for radial timelike geodesics. We can use the same idea to extend this result to non-radial timelike geodesics. For null geodesics we have seen that the condition on $n$ for the divergence of the affine parameter was different for the radial and non-radial cases. But in a timelike case, the angular momentum contribution becomes subleading and we get the same condition for proper time divergence in both cases. So we can conclude that all causal geodesics approaching $r=0$ are complete: timelike geodesics require infinite conformal proper time, while null geodesics require infinite conformal affine parameter to reach $r=0$ singular surface when $n\ge3$. 
%
\section{Conclusion}
In this work, we analyzed the effect of the quantum correction term in the quantum Raychaudhuri equation for radial Klein-Gordon profiles in Schwarzschild spacetime, which governs the quantum corrected geodesic motion of probes. 
Our central result is that quantum probes do not experience the classical Schwarzschild metric directly but instead propagate on a conformally related geometry whose structure is dynamically determined by the quantum state of matter. We then showed that the Schwarzschild black hole singularity can be resolved in a physically meaningful way by incorporating the role of quantum matter in defining an effective spacetime geometry. For the class of Klein-Gordon field configurations considered here, the associated effective geometry becomes curvature regular near the classical singular surface and the corresponding phase flow lines are geodesically complete toward $r=0$.
One outcome of this analysis is that the conformally related metric exhibits finite curvature invariants at the singularity of the Schwarzschild black hole. In particular, the Kretschmann scalar constructed from the dressed geometry remains finite as the radial coordinate approaches zero, despite diverging in the classical Schwarzschild spacetime. Furthermore, with the existence of a smooth interior coordinate system, the apparent divergence of the metric coefficients at $r=0$ does not signal a curvature singularity. Instead, for $n\ge3$, the surface $r=0$ is pushed to infinite geodesic distance and therefore represents an asymptotic boundary of the conformally related spacetime, rather than a removable coordinate singularity.
As a consequence, the proper time or affine parameter required to reach the classical singular location becomes infinite. The dressed spacetime is therefore geodesically complete, and the classical singularity is pushed beyond the physical reach of quantum probes. This conclusion is state dependent and does not imply singularity avoidance for arbitrary quantum states. In the regime where the total quantum and curvature dependent correction is negligible, $\Omega^2\simeq1$, and the ordinary Schwarzschild geodesic dynamics is recovered.

Throughout the analysis, the Schwarzschild metric is treated as a prescribed classical background, and the backreaction of the quantum probe on the spacetime geometry is neglected. This approximation is appropriate within the test field regime considered here, which can be controlled by a suitable rescaling of the overall probe amplitude. In this precise sense, the result is not a claim that the classical Schwarzschild geometry itself is removed, but rather that the admissible class of quantum probes considered here experience an effective geometry in which the classical $r=0$ surface is pushed to infinite effective distance. We also emphasize that vacuum polarization and particle creation are not assumed to be physically absent, they lie outside the approximation scheme adopted here. Such an extension is beyond the scope of the present work and is left for future investigation. Therefore, the conclusion established here is a fixed background Klein-Gordon phase flow statement. Although the classical Schwarzschild spacetime remains geometrically singular for idealized classical probes, the singular surface becomes operationally inaccessible to the quantum probes considered in this paper.

The present framework also suggests several natural extensions. One direction is to apply the same Bohmian conformal construction to cosmological singularities, where the question is whether quantum matter probes in an FLRW background experience the classical big bang surface as an infinite distance boundary in an effective metric. Another important extension is the Kerr geometry, where rotation changes the causal and geometric structure of the singularity. It would also be useful to go beyond the free scalar probes considered here. Interacting scalar fields and fields with spin provide nontrivial extensions of the present construction. These generalizations will test whether the operational regularization found here is a special feature of free Klein-Gordon probes in the Schwarzschild near core geometry or part of a broader mechanism by which relativistic quantum probes render classical singular surfaces physically inaccessible.
\section{Acknowledgment}
This work was supported by the Natural Sciences and Engineering Research Council of Canada. Research fellowships from the Council of Scientific and Industrial Research, India, supported V. Cheriyodathillathu.
\appendix
\renewcommand{\theequation}{A\arabic{equation}}
\setcounter{equation}{0}
\section*{Appendix: Quantifying the test field approximation}
In this appendix we justify the test-field approximation used in the main text. We begin with the Madelung–Bohm decomposition of the Klein–Gordon field and introduce a constant scale factor $\epsilon$ to scale the entire field as follows,
\begin{equation}
    \Phi(x) \longrightarrow \Phi_{\epsilon}(x) = \epsilon \Phi(x)
\end{equation}
Under this transformation, the conformal factor remains unchanged,
\begin{equation}
    \Omega^{2}[\epsilon \Phi] = \Omega^{2} [\Phi]
\end{equation}
Similarly, the transformation of the energy-momentum tensor under the scaling transformation follows directly from its definition in terms of the matter action,
\begin{equation}
T_{\mu\nu}=-\frac{2}{\sqrt{-g}}\frac{\delta S_{\Phi}}{\delta g^{\mu\nu}}
\end{equation}
\begin{equation}
    \Rightarrow T_{\mu \nu}[\epsilon \Phi] = \epsilon^{2} T_{\mu \nu}[\Phi] 
    \label{energy}
\end{equation}
For the above scaling estimate, we only need the fact that the action for the matter field is quadratic in $\Phi$ and in its first derivatives. This means, "$\Omega^{2}$ is invariant under amplitude rescaling, while $T_{\mu \nu}$ is suppressed by $\epsilon^{2}$". We choose an appropriate $\epsilon$ so that we can control the growth of energy density.

For quantifying the "probe regime", we compare the curvature sourced by the probe with the curvature already present in the Schwarzschild background. The test field approximation is valid as long as the curvature scale produced by the probe is much less than the curvature scale already present in the background. For quantifying this limit, we define a backreaction parameter as follows,
\begin{equation}
    \mathcal{B}(r) = \frac{8 \pi G \left\lVert T^{(\Phi)} \right\rVert_{loc}}{\sqrt{\mathcal{K}}}
\end{equation}
$\left\lVert T^{(\Phi)} \right\rVert_{loc}$ is the norm of probe's energy-momentum tensor measured by the freely falling observer in a local orthonormal frame and $\mathcal{K}$ is the Kretschmann scalar associated with the background spacetime. This means, if $ \mathcal{B}(r) <<1$ we can neglect those backreaction effects of probe on the background geometry. From our earlier identity, the back reaction parameter transforms under the field rescaling as follows,
\begin{equation}
    \mathcal{B}_{\epsilon}(r) = \epsilon^{2} \mathcal{B}(r)
\end{equation}
So, we could scale the field with an arbitrarily small $\epsilon$ to be in the valid regime where the backreaction effects are negligible.

Consider one reference solution $\Phi_{0}$ and define its local backreaction parameter as $\mathcal{B}_{1}(r)$. The subscript 1 denotes the unscaled reference field. Because the numerator is a positive local magnitude and the denominator is positive for $r>0$, we have,
\begin{equation}
    \mathcal{B}_{1}(r) \ge 0
\end{equation}
Now we choose an interval $D_{\gamma}=[\gamma, r_{0}]$, where $\gamma$ is the minimum value of $r$. Assuming $\mathcal{B}_{1}(r)$ is continuous on this compact closed interval, it will have a largest value somewhere in the interval by extreme value theorem. We define this maximum value as,
\begin{equation}
    \mathcal{B}_{\max}(\gamma) = \max_{r\in[\gamma,r_0]} \mathcal{B}_{1}(r)
\end{equation}
\begin{equation}
    \Rightarrow \mathcal{B}_{1}(r) \le \mathcal{B}_{\max}(\gamma) \hspace{7pt}\text{for every} \hspace{7pt} r\in[\gamma,r_0]
\end{equation}
Then the rescaled backreaction parameter goes as follows,
\begin{equation}
    \mathcal{B}_{\epsilon_{\gamma}}(r) = \epsilon_{\gamma}^{2} \mathcal{B}_{1}(r) \le \epsilon_{\gamma}^{2} \mathcal{B}_{\max}(\gamma)
    \label{con1}
\end{equation}
For our test field approximation to be valid we need the following condition,
\begin{equation}
    \mathcal{B}_{\epsilon}(r) < \delta \hspace{10pt} \text{for} \hspace{10pt} 0<\delta\ll1
    \label{con2}
\end{equation}
Equations (\ref{con1}) and (\ref{con2}) give us a condition on the choice of the scaling parameter,
\begin{equation}
    \epsilon_{\gamma}^{2} < \frac{\delta}{\mathcal{B}_{\max}(\gamma)}
\end{equation}
One easy choice is,
\begin{equation}
   \epsilon_{\gamma}^{2}  = \min\left\{1,\,\frac{\delta}{2\mathcal{B}_{\max}(\gamma)}\right\}
\end{equation}
If the original probe is already weak, we choose $1$ as the scaling parameter from the above relation. The maximum of backreaction parameter depends on the cutoff,
\begin{equation}
    \mathcal{B}_{\max} = \mathcal{B}_{\max}(\gamma)
\end{equation}
As $\gamma$ moves closer to zero, the interval includes regions of stronger curvature and potentially larger energy densities for the probe. Therefore, typically $\mathcal{B}_{\max}(\gamma)$ increases as $\gamma \to 0$. Consequently, the amplitude required to keep the probe weak must become smaller.

\begin{equation}
    \epsilon_{\gamma}^{2} = \frac{\delta}{2\mathcal{B}_{\max}(\gamma)} \,\,\longrightarrow \,\,0
\end{equation}

The limiting procedure $\gamma \to 0$, $\epsilon_{\gamma} \to 0$ and consequently $\mathcal{B}_{\epsilon_{\gamma}}(r) < \delta$, should be interpreted as a sequence of progressively weaker probe fields defined on progressively deeper domains, rather than as a single solution with a continuously varying normalization.

In summary, the effective geometry is independent of the overall probe amplitude, while backreaction is amplitude suppressed. So, we consider a family of weaker and weaker probes that explores smaller and smaller radii. Since $\Omega^{2}$ is unchanged under $\Phi \to \epsilon \Phi$, the effective geometry is identical for all the members of the family. In contrast, the energy momentum tensor scales quadratically ($T_{\mu \nu}[\epsilon \Phi] = \epsilon^{2} T_{\mu \nu}[\Phi]$), leading to a corresponding suppression of the backreaction. Therefore, as the cutoff surface is moved inward, the parameter $\epsilon$ is simultaneously reduced so that the backreaction parameter satisfies, $\mathcal{B}_{\epsilon_{\gamma}}(r) < \delta$, throughout the region of interest. This constitutes the basis of the test-field approximation employed in our analysis.

\nocite{PhysRev.98.1123}
\nocite{PhysRevD.89.084068}
\nocite{HawkingPenrose1970}
\nocite{Penrose1965}
\nocite{Polchinski1998}
\nocite{Zwiebach2009}
\nocite{Rovelli2004}
\nocite{AshtekarLewandowski2004}
\nocite{Reuter1998}
\nocite{Percacci2009}
\nocite{PhysRev.85.166}
\nocite{doi:10.1142/S0217751X98000305}
\nocite{Broglie1960NonlinearWM}
\nocite{Chowdhury:2021fjw}
\nocite{Tangherlini:1963bw}
\nocite{Singh:2017vfr}
\nocite{Wald:1984rg}
\nocite{Nikolic:2004tu}
\nocite{Alkhateeb:2022vcf}
\nocite{Shojai:1998qou}
\nocite{article}
\nocite{ALI2016173}
\nocite{doi:10.1142/S0217732319502705}
\nocite{PhysRevLett.93.090402}
\nocite{10.1143/ptp/90.4.753}
\nocite{GeorgeHorton_2000}
\nocite{10.1098/rspa.2013.0699}
\nocite{HrvojeNikolic2007}
\nocite{EricGreenwood_2008}
\nocite{PhysRevD.89.044003}
\nocite{PhysRevD.52.5670}
\nocite{Foo_2022}
\nocite{PhysRevA.111.052201}
\nocite{hatifi2025quantumgravitymetricquantization}
\nocite{LANTIGUA2025170226}
\nocite{Bilaniuk1962}
\nocite{Bilaniuk1969}
\bibliographystyle{ieeetr}
\bibliography{Revised-Manuscript}

\end{document}